\documentclass{elsart}

\usepackage{epsfig, upgreek}

\usepackage{amssymb}

\begin{document}

\begin{frontmatter}



\title{Simulation of radio emission from air showers in atmospheric electric fields}
\author[nijmegen,lbnl]{S. Buitink\corauthref{cor1}}
\ead{sbuitink@lbl.gov},
\author[fzk]{T. Huege},
\author[nijmegen]{H. Falcke},
\author[nijmegen]{J. Kuijpers}
\corauth[cor1]{Corresponding author.}
\address[nijmegen]{Department of Astrophysics/IMAPP, Radboud University Nijmegen, P.O. Box 9010, 
6500 GL Nijmegen, The Netherlands}
\address[lbnl]{Lawrence Berkeley National Laboratory, Berkeley, California 94720}
\address[fzk]{Institut f\"ur Kernphysik, Karlsruher Institut f\"ur Technologie, Postfach 3640, 76021 Karlsruhe, Germany}

\begin{abstract}
We study the effect of atmospheric electric fields on the radio pulse emitted by cosmic ray air showers. Under fair weather conditions the dominant part of the radio emission is
driven by the geomagnetic field. When the shower charges are accelerated and deflected in an electric field
additional radiation is emitted. We simulate this effect with the Monte Carlo code REAS2, using CORSIKA-simulated showers as input. In both codes a routine has been implemented
that treats the effect of the electric field on the shower particles. We find that the radio pulse is significantly altered in background fields of the order of $\sim 100$~V/cm 
and
higher. Practically, this means that air showers passing through thunderstorms emit radio pulses that are not a reliable measure for the shower energy. Under other weather 
circumstances
significant electric field effects are expected to occur rarely, but nimbostratus clouds can harbor fields that are large enough. In general, the contribution of the 
electric field to the radio pulse has polarization properties that are
different from the geomagnetic pulse. In order to filter out radio pulses that have been affected by electric field effects, radio air shower experiments should keep weather
information and perform full polarization measurements of the radio signal.  
\end{abstract}


\end{frontmatter}

\section{Introduction}
\label{sec:intro}
In recent years the technique of radio detection of cosmic ray air showers has developed into a promising
detection mechanism. Plans for LOFAR \cite{lofar} triggered renewed interest \cite{FG03} in this technique that was first 
explored in the 1970s.
With LOPES, the LOFAR prototype station, it has been established that the radio pulse power is proportional to the square
of shower energy and dependent on the angle of the shower axis with the Earth's magnetic field \cite{F05}. These results show
that the main emission mechanism is coherent and driven by the geomagnetic field. The good angular resolution and high duty cycle of radio antennas make them an attractive addition to large air shower arrays like the
Pierre Auger Observatory \cite{PAO}.

The emission mechanism can be described both microscopically, as coherent synchrotron emission from the shower
electrons and positrons that follow curved trajectories in the magnetic field, and macroscopically, as radiation from a transverse current that develops as the shower charges are driven apart by the magnetic field. The
first approach was proposed by Falcke \& Gorham \cite{FG03} and worked out in thorough detail in Huege \& Falcke \cite{HF03,HF05}. The
second approach was first described by Kahn \& Lerche \cite{KL66} and has recently been improved by Scholten et al.\ \cite{Sch} and Werner \& Scholten \cite{ws07}. 

Already in the 1970s it was discovered that the radio pulse of an air shower may be larger than anticipated when
strong electric fields are present in the atmosphere \cite{M74}. Using LOPES data recorded during various weather types it was
shown that this amplification of the radio pulse only occurs during thunderstorm conditions \cite{B07}. In another study it was
shown that the arrival direction reconstructed with radio data and particle detector data can differ by a few 
degrees during thunderstorms \cite{N08}.

To further explore the nature of the electric field effect on air showers and the conditions under which it becomes
important, CORSIKA simulations were carried out \cite{B09}. These simulations showed that shower electron and positron energy 
distributions can strongly be affected in background electric fields of the order of $\sim 1$~kV/cm. When the electric field strength
exceeds a certain threshold \cite{D05} electron runaway breakdown is observed leading to an exponential increase in the number of
electrons, an effect first predicted by Gurevich et al.~\cite{G92}. 

In this work we simulate the effect of electric fields on the strength of the radio pulse with REAS2 using the results
from Buitink et al.\ \cite{B09} as input. Not only can the altered energy distributions of shower particles influence the radio signal, also
the emission mechanism itself changes, as we will describe in the next Section.
  
\section{Emission mechanism}
\label{sec:effects}
To understand the effect of electric fields on the radiation of air showers we first consider the basic emission theory. 
Currently, two emission mechanisms are advocated: the geosynchrotron model and the transverse current model. The former is a microscopic model while the latter is a macroscopic model. They are similar at first glance but contain some important
differences that will have to be resolved in the future.
 
In the Lorentz gauge the vector potential for a moving particle with charge $q$ and velocity ${\bf \upbeta}={\bf v}/c$ 
can be expressed in the Li\'enard-Wiechart form (Eqn.~14.8 in Jackson 1975):
\begin{equation}
\label{lw_p}
{\bf A}({\bf r},t)=\left. \frac{q{\bf \upbeta}}{(1-{\mathbf \upbeta}\cdot{\bf n}) R}\right|_{\mathrm{ret}},
\end{equation}
where ${\bf n}$ is the unit vector from the charge in the direction of the observer and $R$ the distance to the observer. The subscript `ret'
indicates that the expression is evaluated at the retarded time. 

For a radiating system of many moving particles, such as
an air shower, two approaches can be chosen. In the macroscopic approach followed by Scholten et al.\ \cite{Sch} the first step is to
write the vector potential of the current produced by all particles together. The contribution of one particle to the
current density is
\begin{equation}
{\bf J}({\bf x},t)=q{\bf v} \delta({\bf x}-{\bf x}^{\prime}),
\end{equation}
where ${\bf x}^{\prime}$ is the location of the particle, leading to
\begin{equation}
\label{lw_j}
{\bf A}({\bf r},t)=\frac{1}{c}\int \left. \frac{{\bf J}({\bf x},t)}{(1-{\bf \upbeta}\cdot{\bf n}) R} \mathrm{d}^3{\bf x} \right|_{\mathrm{ret}},
\end{equation}
where $\upbeta$ is now understood as the velocity of the current element.  

The total current density is found by summing
over all particles. If we assume a charge neutral shower, the current density in the direction parallel to the shower
axis vanishes. The mean transverse drift caused by the magnetic field is opposite and equal for electrons and
positrons and so the current density can be written as:

\begin{equation}
\label{tcurrent}
{\bf J}({\bf x},t)=n({\bf x},t) \langle q{ \bf v}_{\mathrm D} \rangle,
\end{equation}
where $n$ is the number density of electrons and positrons, $q$ is the charge of a particle, ${ \bf v}_{\mathrm D}$ its transverse drift velocity, and the brackets denote the mean value over all particles. The distribution $n({\bf x},t)$ describes the shape and evolution of the shower front, and the total number of charges in the shower front at a certain moment, and can be approximated by using shower parametrizations.

A simple approximation for an air shower is one in which all particles are at the same point, so 
\begin{equation}
{\bf J}({\bf x},t)= N(t) \langle q{\bf v_D} \rangle \delta({\bf x}-{\bf x_i}^{\prime}),
\end{equation}
where $N(t)$ is the number of particles at time $t$ and ${\bf x}^{\prime}$ is the location containing all the particles.
The radiation field is found by taking the time derivative of Eqn.\ \ref{lw_j}
and it can be seen immediately that this produces a bipolar pulse with $E \propto \mathrm{d}N/\mathrm{d}t$. A more realistic distribution of charges will conserve the bipolar character of the pulse \cite{Sch}. 

In this approach it is not evident whether it is justified to take the mean value of the drift velocity before taking the
time derivative. The choice to disregard the acceleration of individual particles is based on the notion that coherent emission can be described by the collective behavior of all particles. The
small-scale acceleration and deceleration of individual charges would then average out to a large-scale constant drift velocity. 
If, however, the velocity would have remained in the equation as a time dependent quantity for the individual
particles a time derivative would have given an extra term which is proportional to $N$. Without this additional term, a shower with 
a constant number of particles would have a constant current density everywhere and no radiation is produced (in sharp contrast to the geosynchrotron approach). 

In the microscopic approach the radiation field is seen as the superposition of the fields of all shower particles.
Taking the time derivative of Eqn.\ \ref{lw_p} we arrive at the radiation equation (Eqn. 14.14 in \cite{Jackson}):
\begin{equation}
\label{radfield}
{\bf E}({\bf x},t)=
e\left[\frac{{\bf n}-{\bf \upbeta}}{\gamma^2(1-{\bf \upbeta}\cdot{\bf n})^3 R^2} \right]_{\mathrm{ret}}
+\frac{e}{c}\left[ \frac{{\bf n}\times\left[ ({\bf n}-{\bf \upbeta})\times{\bf \dot{\upbeta}}
\right] }{(1-{\bf \upbeta}\cdot{\bf n})^{3}R}\right]_{\mathrm{ret}}.
\end{equation}

For the acceleration of a charge in a magnetic field this equation will yield synchrotron radiation. The air shower radio simulation code
REAS2 \cite{H07} calculates the synchrotron contribution for a representative part of the trajectories of all shower charges.
By summing the synchrotron contributions of all particles, a radio pulse is found that is unipolar 
and is roughly proportional to $N(t)$. 
In this approach the summation over all particles is performed after taking the time derivative of the Li\'enard-Wiechart potential. 
However, if we write the radiation field as:
\begin{equation}
{\bf E} = \frac{\mathrm{d}}{\mathrm{d}t} \sum_i^{N(t)} {\bf A}_i,
\end{equation}
it is clear that the summation can not be put in front of the time derivative because the number of shower particles $N$ is
dependent on $t$.  A correct time derivative would produce an additional term proportional to $\mathrm{d}N/\mathrm{d}t$, and can be associated to the growth and decay of a
transverse current. 

A way to work around this problem is to keep the number of particles constant. We can think of the
charges as being at rest since $t=-\infty$ and quickly being accelerated to their velocity $\beta$ at the time of their
creation. At the end of its track the charge is decelerated again until it is at rest where it remains until
$t=\infty$. These periods of acceleration and deceleration give rise to additional radiation, which is currently not included in REAS2.

A way to work around this problem is to keep the number of particles constant. In the shower, electrons and positrons are constantly created
in pairs. For calculating the radiation one might as well think of these
charges as being at rest since $t=-\infty$ and quickly being accelerated to their velocity $\upbeta$ at the time of their
creation. Since the electron and positron are co-located at the moment of creation, there is no charge density present until the charges are
separated in the magnetic field. Hence, no radiation is produced in the creation or `acceleration' of the charges. 

At the same time, particles should be traced until they have lost most of their energy and they can be regarded as stationary\footnote{Note that these particles still produce a
stationary field. This field corresponds to the dipole field left behind by the shower in the atmosphere.}. 
The charges decelerate because of various processes such as radiation losses, bremsstrahlung and
positron annihilation. In such processes the momentum vector of the charge will change and there will be electromagnetic radiation. The
polarity of these \emph{deceleration pulses} is determined by the sign of the charge, and for an equal amount of electrons and positrons
in the shower, the pulses will add up incoherently, \emph{as long as} the momentum distributions of electrons does not differ from that of positrons.

However, since the the shower electrons and positrons are separated in the magnetic field they gain an opposite transverse momentum, for
which the deceleration radiation adds up coherently. There will be a coherent deceleration radiation component that is driven by the magnetic
field. Although deceleration processes can be interactions that happen on very short time scales, the
wavelengths at which the deceleration emission is coherent is given by the size of the shower front, as is the case for geosynchrotron
emission.
The processes of geosynchrotron emission and coherent deceleration emission are closely related, since both are driven by the magnetic field.
Inclusion of the contribution from deceleration into REAS2 is planned and currently under investigation.

Furthermore, the calculations in REAS2 are based on an index of refraction of
unity. Therefore, contributions to the radio emission by Cherenkov or transition radiation are not included, nor the influence of the medium
on the propagation of the radio waves.

In this work we use the REAS2 code. When no electric field is present in the atmosphere the
trajectory of a charge can be described as a gyration in the magnetic field. Inserting the parameters of this motion into the radiation
equation \ref{radfield} gives synchrotron radiation. When both an electric and a magnetic field are present in the atmosphere the charges
follow a more complicated trajectory that is derived in Appendix \ref{app:trajectory}. When the parameters of this type of motion are
inserted in the radiation equation, a radiation field is found that is different from pure synchrotron radiation and includes the radiation of
acceleration in the electric field.

\section{Simulation setup}
\label{sec:setup}
REAS2 \cite{H07} is a Monte Carlo code that calculates the geosynchrotron emission from air showers that are simulated with CORSIKA
\cite{corsika}. The electromagnetic interactions in CORSIKA are based on the EGS4 code \cite{EGS4}, which includes all possible interactions (including elastic scattering and proper treatment of multiple scattering and ionization energy loss).
In order to correctly implement the effect of a background electric field, an electric field routine is implemented in both
REAS2 and CORSIKA. With the modified version of CORSIKA we have shown that a background field strength of 1000~V/cm in order of magnitude can
significantly change the energy distribution of the electrons and positrons. Also, high in the atmosphere, where the air density
is low enough, avalanches of runaway electrons can occur. A detailed description of the results of our CORSIKA simulations is presented in Buitink et al.\ \cite{B09}.

We used the COAST plugin \cite{coast} to output particle distributions at 50 layers in the atmosphere. For each layer two
three-dimensional histograms
are produced. One containing:
\begin{itemize}
\item particle energy,
\item distance of the particle to the shower axis and,
\item delay time of the particle with respect to a (virtual) shower front that travels with the speed of light.
\end{itemize}
The second distribution contains:
\begin{itemize}
\item particle energy,
\item angle between the particle momentum and shower axis and,
\item angle between the component of the particle momentum that is perpendicular to the shower axis and a vector pointing radially outwards
from the shower axis to the particle.
\end{itemize}
Both histograms are created separately for electron and positrons.  

From these distributions REAS2 picks particles and follows a small part of their trajectories. In order to do this, an analytic
expression for the particle trajectory has to be implemented which gives the particle momentum and acceleration at
various points of the trajectory. The electric field effect is included in REAS2 by implementing the equations of
motion for a charge inside a homogeneous electric and magnetic field which are under some angle. These expressions are
derived in Appendix \ref{app:trajectory}. The radiation at ground level for observers at various locations is calculated by adding the contributions of Eqn.
\ref{radfield} for a representative amount of shower particles.

We use a shower with a proton of $10^{16}$~eV as a primary particle. Simulations are done for a vertical shower and showers with a
$30^{\circ}$ and $60^{\circ}$ inclination angle. The inclined showers are simulated in CORSIKA as moving towards the north. These showers
have been rotated in REAS2 to calculate the emission of showers from other arrival directions. Although there is certainly a difference between CORSIKA simulations of showers
propagating in different directions, the histograms do not capture these azimuthal asymmetries, and including the correct azimuth angle in the CORSIKA simulation will therefore
not give more precise results.

The electric fields that are used are vertically aligned and have strengths of 100 and 1000 V/cm. The largest field strengths that have been found in
thunderstorms are of the latter order of magnitude \cite{book}. To gain a good understanding of the effect we use a homogeneous electric field.
Inside thunderstorms the strength and polarity of the field will vary with altitude. The implications of our results for more realistic field
configurations is discussed in Section \ref{sec:discussion}.
  
REAS2 currently calculates only the synchrotron contribution of the shower electrons and positrons.
In some of our CORSIKA showers that were simulated in a large electric field background an exponential growth in the
number of electrons is observed. The radio emission that is associated with the growth and decay of the vertical
current that is produced in this way is not simulated by REAS2 (see Section \ref{sec:effects}). The radio emission of this component has been the subject of several studies \cite{G02,T05,D09} and its characteristics will be described in the end of the next Section.

\section{Simulation results}
\label{sec:results}
We present the full polarization information of the pulses simulated with the REAS2 code including electric field effects. This is done by plotting the three spatial
components of the radiation field at the location of the observer, which we call East-West (EW), North-South (NS) and vertical or $z$ polarization. In principle the polarization of a
pulse can be fully described with two components perpendicular to the direction of propagation of the radio wave. The three-dimensional representation is chosen because in a
typical air shower radio experiment, antennas are not aimed at the direction from which the radiation is coming. Instead, the antennas are mounted in fixed positions, for which
EW, NS, and $z$ alignments are the most straightforward choices. A second reason for the three-dimensional representation is that the direction of propagation of the radio wave
cannot be unambiguously defined for extended radio sourced close to the receiver. Although the bulk of the radiation is expected to come from the direction of the region around 
the shower maximum, all parts of the shower contribute to the observed pulse. Depending on the the shower geometry, these contributions may come from different directions.   

The electric field is defined in such a way that a positive field points downwards. In other words: in a positive field, positrons are accelerated downwards 
and electrons are accelerated upwards.

\begin{figure}[htp]
\centerline{\epsfig{file=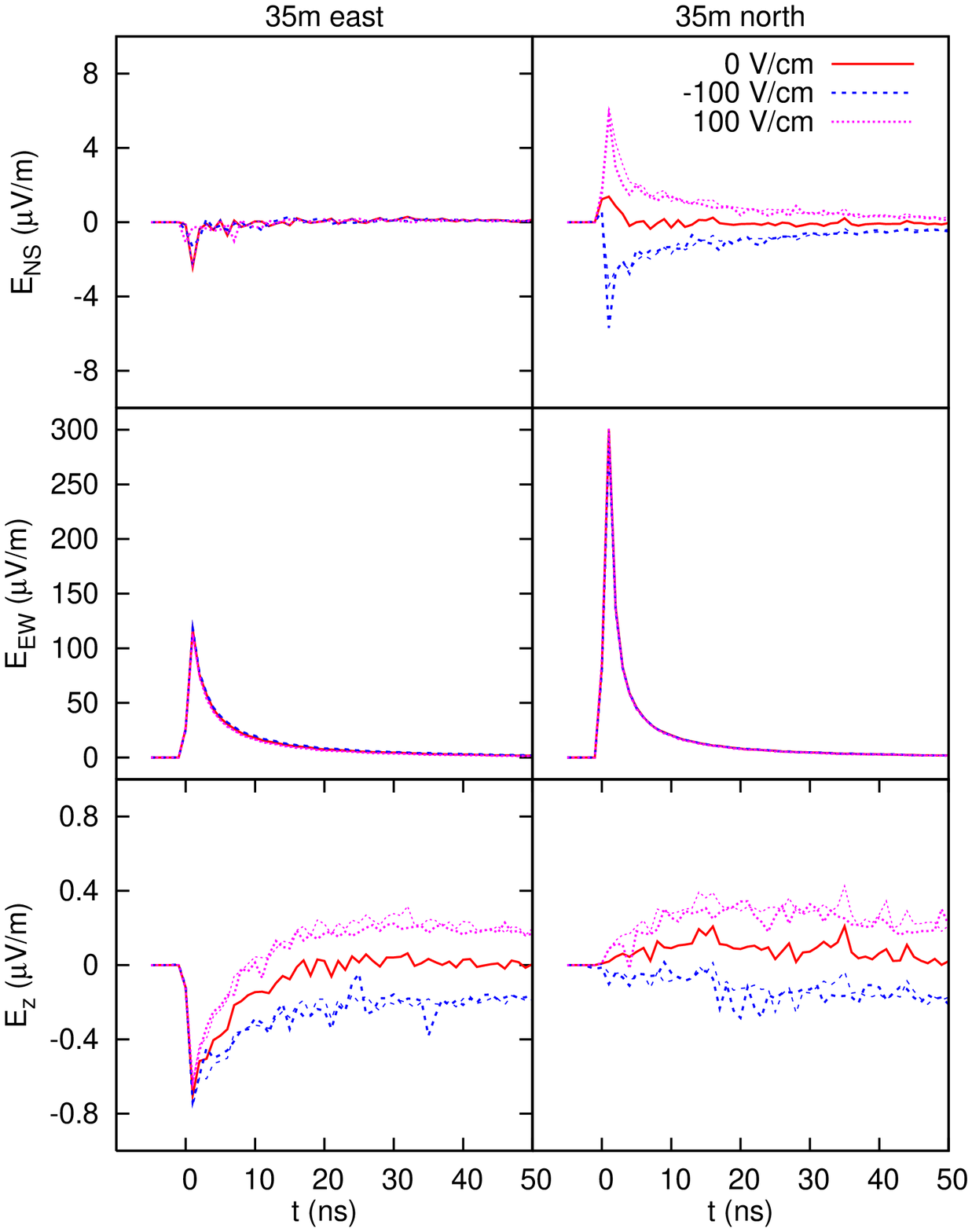 ,width=0.5\textwidth}
            \epsfig{file=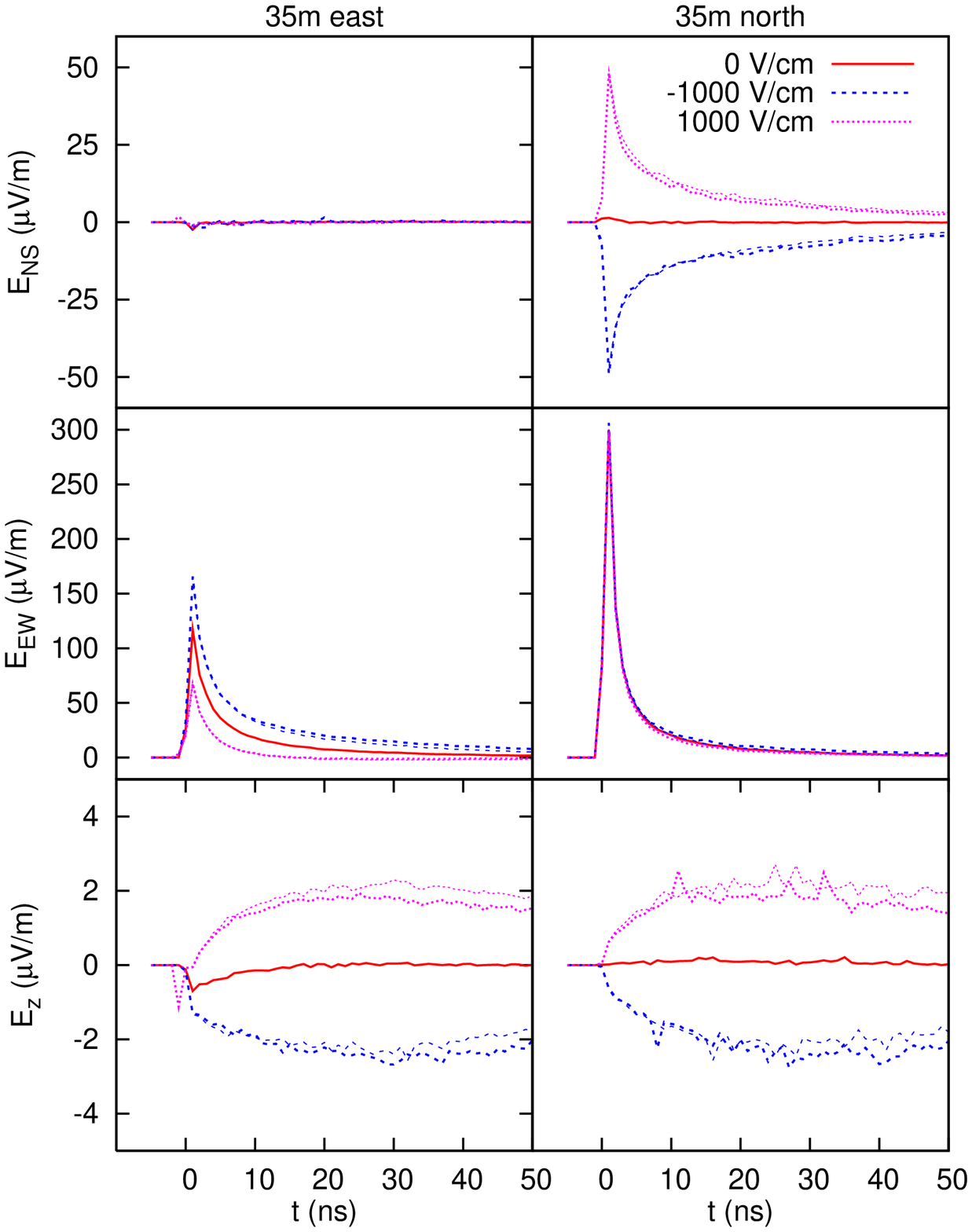 ,width=0.5\textwidth}}
\caption{Radio pulses for a vertical shower of 10$^{16}$ eV in the presence of electric fields of 0, 100 V/cm (left panel), and 1000 V/cm (right panel). Pulses in the NS, EW and $z$
polarization are shown for observers located 35~m to the east and the north of the shower core. Thick lines correspond to simulations in which the electric field effect is
switched on in both CORSIKA and REAS2. Thin lines (almost indistinguishable in this case) correspond to simulations in which the electric field routine is only switched on in
REAS2. Note that the scales on the y-axes of different polarizations is different.}
\label{multivert35}
\end{figure}
In Fig. \ref{multivert35} the simulated pulses are plotted for two observers that are located respectively 35~m east and north from the
shower core for a vertical shower of $10^{16}$~eV in background electric fields of 100 and 1000~V/cm. For each pulse all
polarizations are plotted. Note that the scales on the y-axes of different polarization is different. It can be seen that the contribution of the electric field to the radiation is dependent on the
location of the observer. To understand this effect, let us consider the polarization of the radiation from a single
particle in an electromagnetic field. In Appendix \ref{app:polarization} we use the assumption that the vector potential
is proportional to the perpendicular part of the total force that works on the particle to arrive
at Eqn.\ \ref{polarization}. This equation can be used to explain the polarization properties observed in the simulations.

Because the radiation from the particles is strongly beamed in the forward direction, observers will predominantly see the
radiation from particles moving towards them. An observer that is situated north of the shower core will mainly see radiation
from particles moving to the north. For these particles $\phi=0$ and $\theta$ can have various angles, depending on the
altitude of the particle. By inserting the magnetic angle $\eta_{B}=26^{\circ}$ (for central Europe, see Eqn.\ \ref{eq:bfield}) we see that in absence of an
electric field, all contributions to the radiation will be linearly polarized in the EW direction. For the same observer, the
contribution of the electric field will show up in the NS and $z$ polarization (mainly NS for $\theta$
close to 0). For an observer east to the shower core ($\phi=270^{\circ}$) the largest component of the synchrotron radiation is 
in the EW plane, but radiation is also observed in the other polarizations. The electric field is affecting the
radiation in the EW and the $z$ plane. All these features are observed in Fig.\ \ref{multivert35}. 

\begin{figure}[htp]
\centerline{\epsfig{file=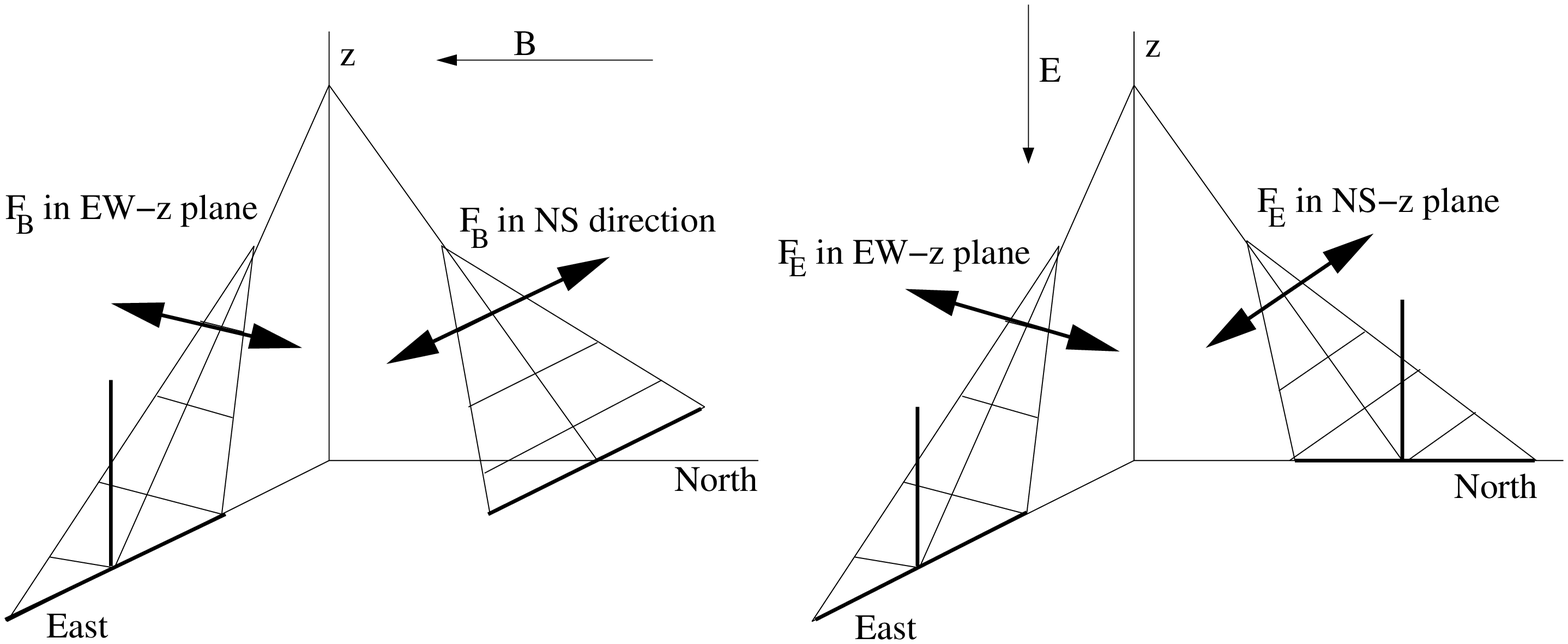 ,width=0.8\textwidth}}
\caption{Schematic view of direction of the electric and magnetic forces and their influence on the polarization properties. See text for details. \label{polofrad}}
\end{figure}

Fig.~\ref{polofrad} illustrates why the polarization properties are different for acceleration in a magnetic and in an electric field. The left picture shows the direction of
the Lorentz force for particle pairs moving in different directions. For simplicity, the magnetic field is chosen to be horizontal and in the NS direction. Particles moving in
the eastern direction will be deflected perpendicular to the direction of propagation in the EW-$z$ plane. An observer on the EW axis will measure a pulse polarized in that
plane. Particles moving towards the north will be deflected in the EW direction. Observers on this axis measure a pulse with a polarization parallel to the EW axis. 

In the right picture there is an electric field in the $z$ direction (and no magnetic field). The electric force generally has a component along the direction of propagation and perpendicular to the
direction of propagation. For the radiation we ignore the former component. The perpendicular component lies in the EW-$z$ plane for particles moving towards the east. An
observer on the EW axis will measure a pulse polarized in the EW-$z$ plane, just like pulses generated in a magnetic field. An observer on the NS axis, however, will see a pulse
that is polarized in the NS-$z$ plane, in contrast to the magnetically generated pulse that is polarized in the EW direction. For this observer, the contributions of a pulse 
that is produced with both an electric and magnetic background field, appear in different polarization directions. 

For an inclined magnetic field and observers located off-axis the polarization properties are more complicated and are described by Eqn.~\ref{polarization}. In general, the
polarization of the magnetic and electric contribution is different.

\begin{figure}[htp]
\centerline{\epsfig{file=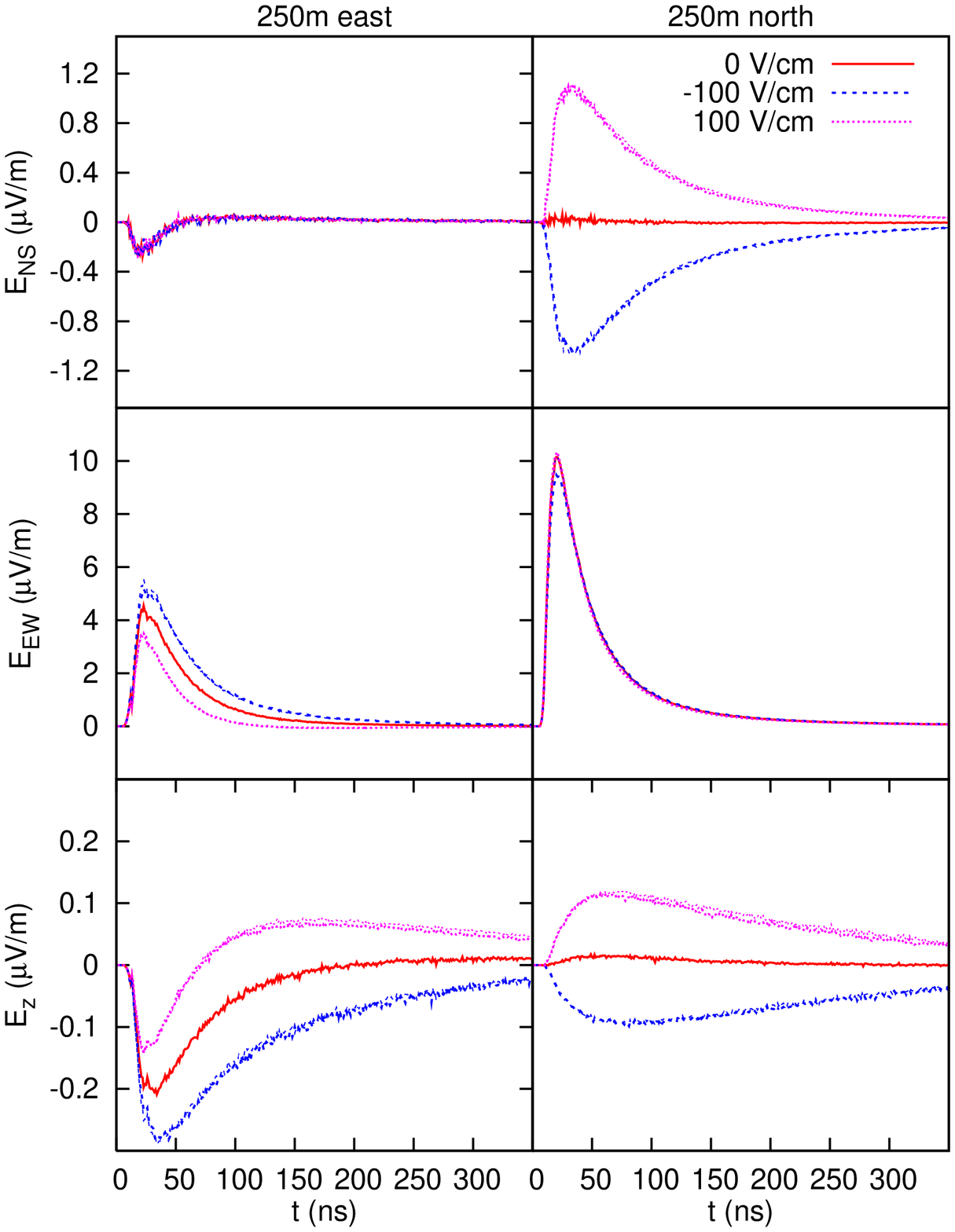 ,width=0.5\textwidth}
            \epsfig{file=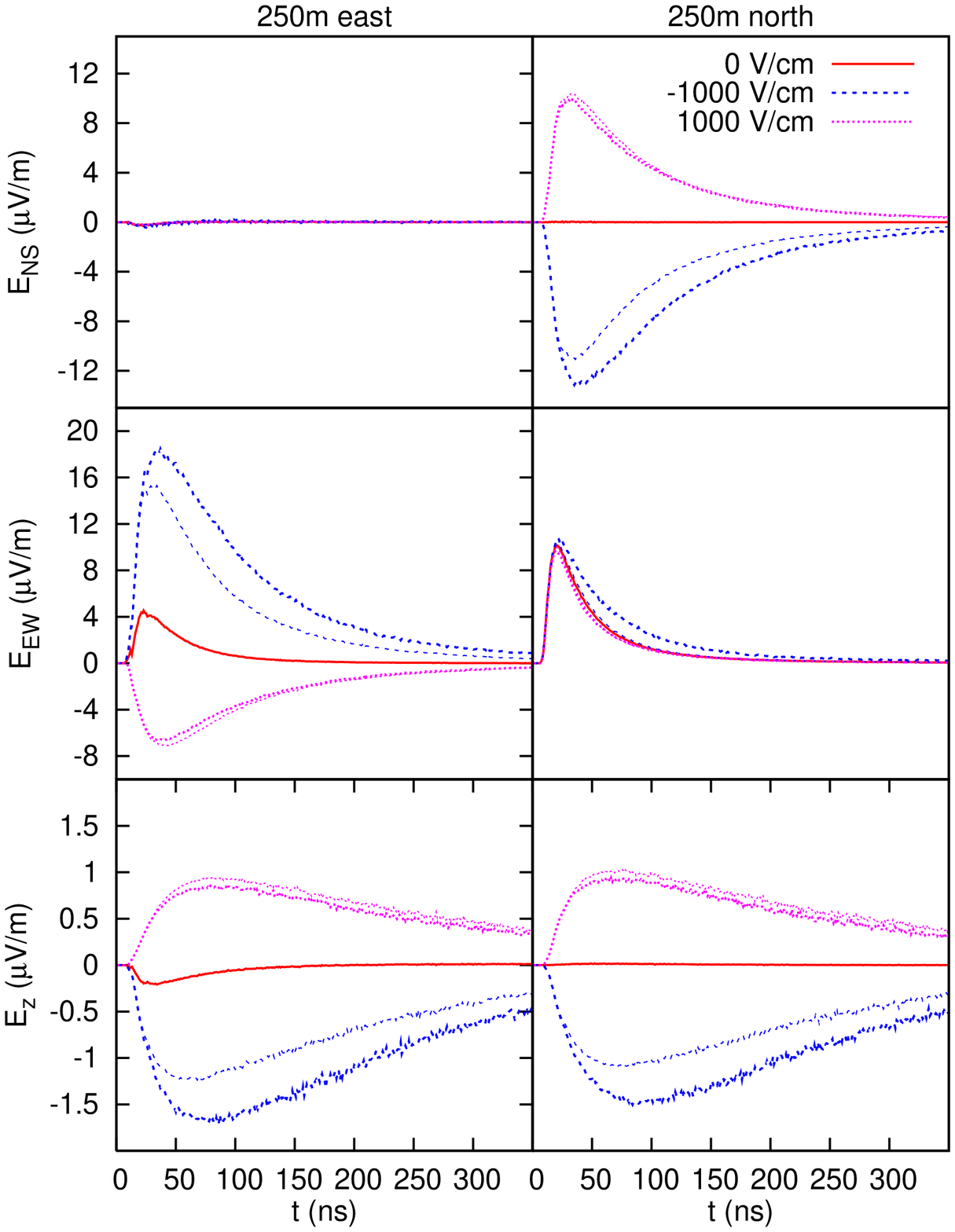 ,width=0.5\textwidth}}
\caption{Same as Fig. \ref{multivert35} but for observers at 250~m from the shower core.}
\label{multivert250}
\end{figure}

The effect of a vertical electric field of 100 V/cm on a vertical shower is almost undetectable for an observer close to the
shower core, but at larger distances its effects become important. Fig.~\ref{multivert250} shows that at 250~m distance, the pulse height variation in the EW
polarization due to the electric field can vary from 0\% (observer in the north) to roughly 25\% (observer in the east). Electric field of the order of 10 V/cm do not give a
significant contribution to the pulse height, not even at larger distances.

For a field of 1000 V/cm an observer to the east can observe variations of $\sim 50$\% in the EW plane already close the shower
core (Fig.\ \ref{multivert35}), while for distant observers the radiation can be enhanced by a factor 4 and even change polarity. Towards the north
the radiation in the EW plane remains unchanged. Instead, the electric field contribution is visible as a pulse in the NS
plane. The polarity of the pulse is dependent on the direction of the electric field, as can be understood from Eqn.\ \ref{polarization}.
When the contribution of the electric field in one of the polarizations has a sign opposite to the magnetic field contribution, the result
can either be a suppressed pulse, or a pulse that has changed polarity. For example, the pulse in in the EW plane for an observer 250~m east
of the shower core is suppressed when the electric field strength is 100 V/cm (see left panel of Fig.\ \ref{multivert250}). When the field
strength is 1000 V/cm the pulse has reversed polarity (see left panel of Fig.\ \ref{multivert250}). In effect, the total intensity of the radio pulse can also increase or
decrease depending on the polarity and strength of the electric field.

By measuring the polarization properties of an air shower that has propagated through an electric field, it is in principle possible to determine the polarity of the field. It
may even be feasible to make an estimate of the field strength at the region where most radiation is emitted, which is at the altitude where the shower reaches its maximum and the region
above it \cite{H07}. 

For each pulse that is simulated in a background electric field, a thick line represents a full simulation in which the
field effects are switched on in both CORSIKA and REAS2, while the thin line represents a simulation in which the field
effects are switched off in CORSIKA. For vertical showers the difference between these lines is hardly visible, indicating that
the change in energy distribution of the particles is not very important for the radiation. 

The only noticeable exception for vertical showers exists for a negatively aligned field of 1000 V/cm (accelerating the electrons, see right panels of Figs.~\ref{multivert250}). For this field configuration 
the number of electrons in the upper atmosphere
has increased explosively due to electron runaway breakdown \cite{B09}. These particles give a contribution that can make 
the radio pulse higher or broader.

A vertical shower in a vertical electric field is a special case in the sense that observers, who are located at different directions from the shower
core, see particles that are moving in different azimuthal directions. For inclined showers, however, all particles move
in roughly the same azimuthal direction and the effects of a vertical electric field will only change slightly between observers with different directions to the shower core. 

\begin{figure}[htp]
\centerline{\epsfig{file=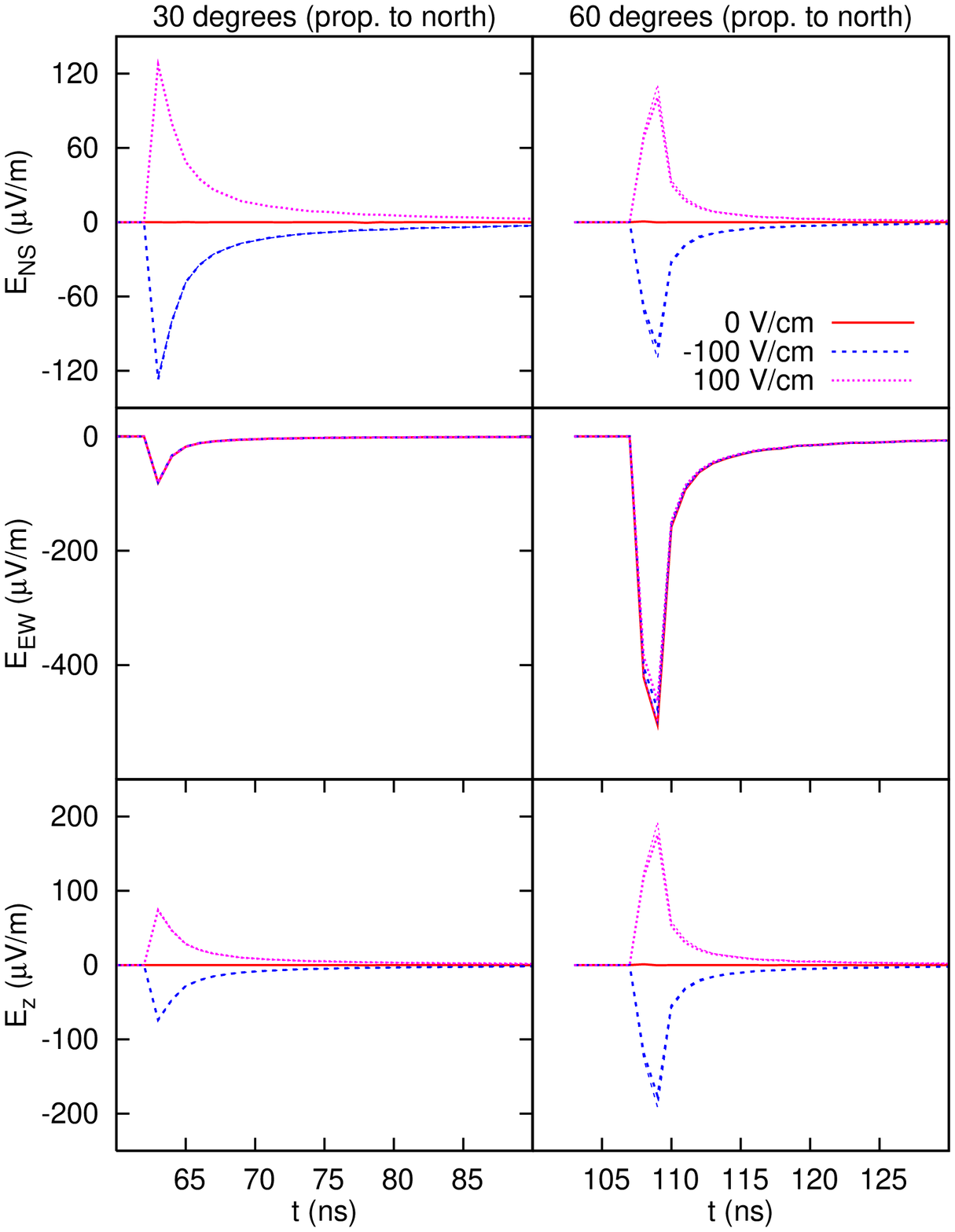 ,width=0.5\textwidth}
            \epsfig{file=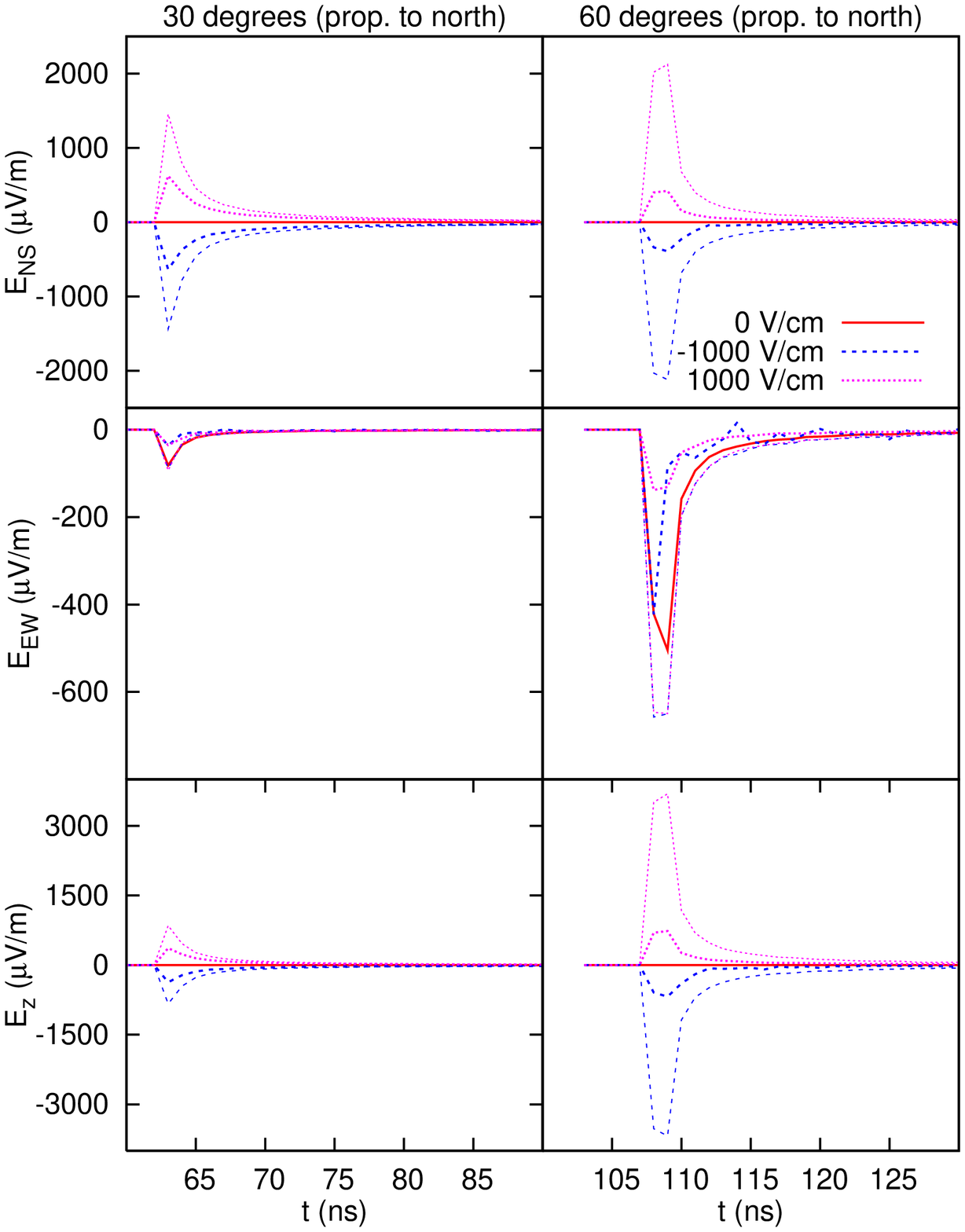 ,width=0.5\textwidth}}
\caption{Radio pulses for inclined showers of 10$^{16}$ eV propagating towards the north in the presence of electric fields of 100 V/cm 
(left panel) and 1000 V/cm (right panel). The polarization of pulses in the NS, EW and $z$ directions
are shown for an observer located 35~m north of the shower core. Thick lines correspond to simulations in which the electric field effect is
switched on in both CORSIKA and REAS2. Thin lines correspond to simulations in which the electric field routine is only switched on in
REAS2.}
\label{multi_az0}
\end{figure}

\begin{figure}[htp]
\centerline{\epsfig{file=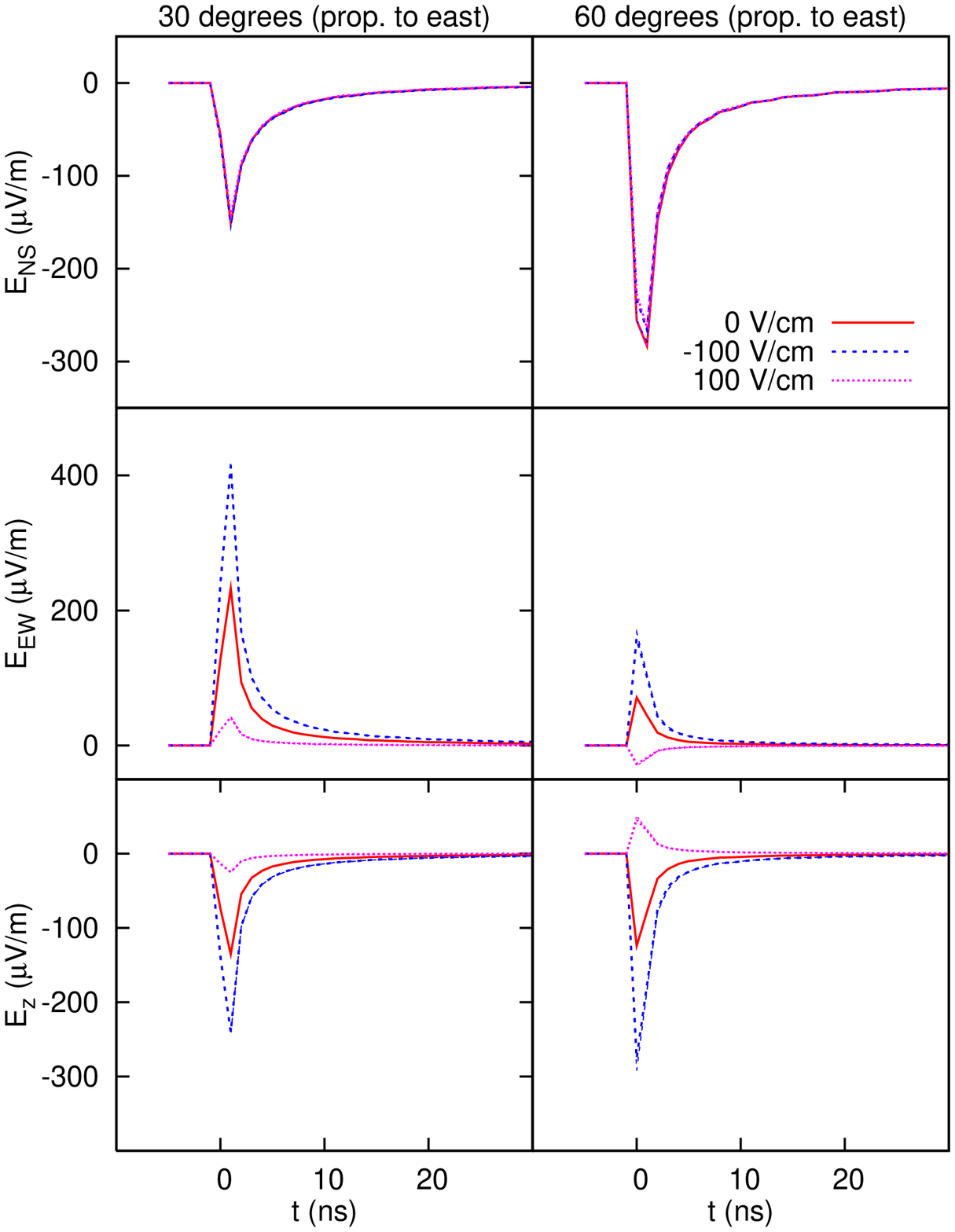 ,width=0.5\textwidth}
            \epsfig{file=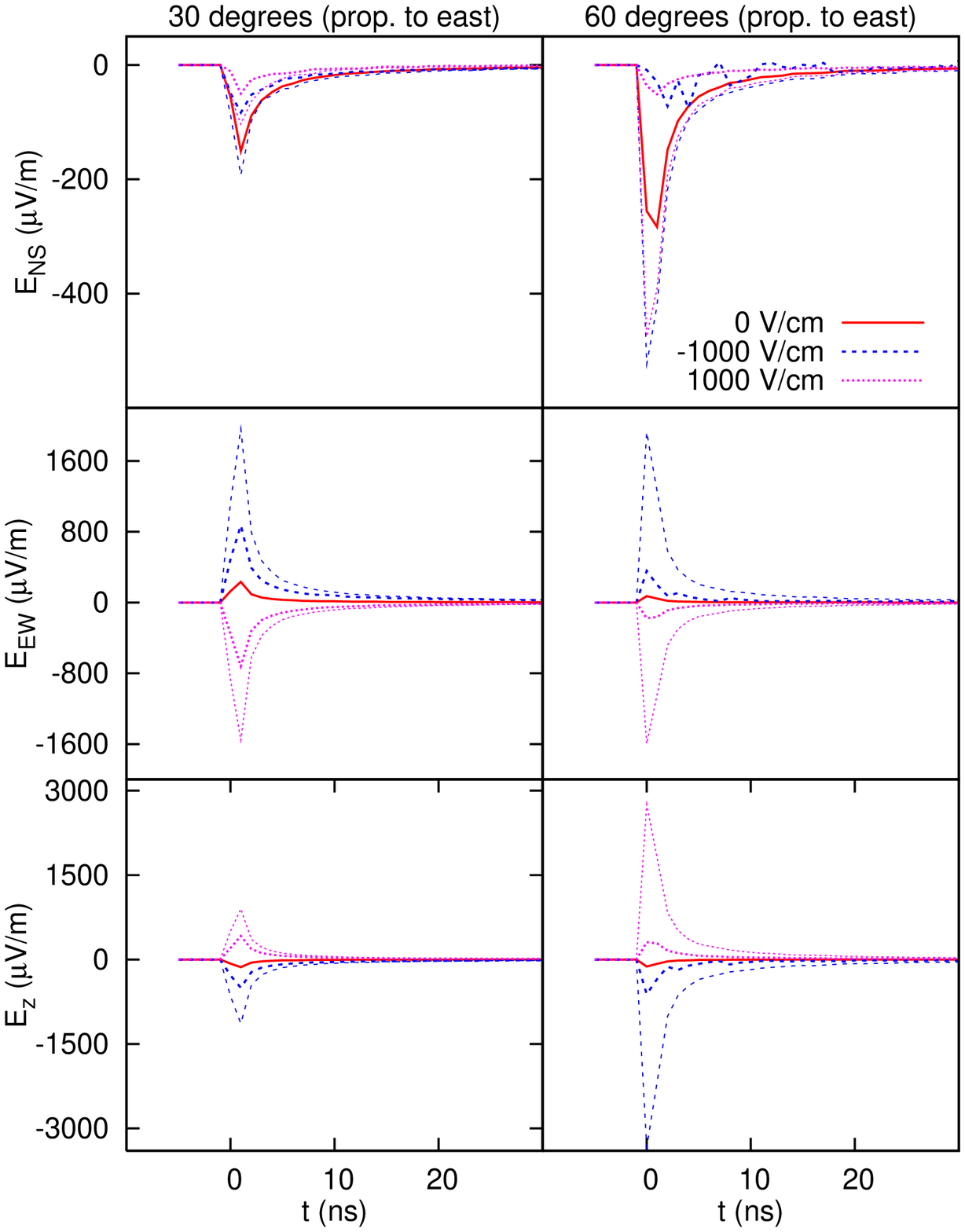 ,width=0.5\textwidth}}
\caption{Same as Fig. \ref{multi_az0} but for showers propagating towards the east.}
\label{multi_az270}
\end{figure}

Fig. \ref{multi_az0} shows radio pulses for an observer 35~m north of the shower core for showers with zenith angles of 30 and 60 degrees propagating towards the north. Note
that this distance is measured horizontally and does not coincide with the distance of the observer to the shower axis. Observers to
the east of the shower core observe similar polarization features as observers to the north, because they also observe particles 
moving roughly towards the north (in contrast to the situation for vertical showers). Indeed, the polarization features are the
same as for the northern observer in the case of a vertical shower. An exception is the polarity of the pulse in the EW plane, which has flipped because the outer product
between the shower direction and the magnetic field has changed sign. The shower with 30$^{\circ}$ zenith angle propagates nearly parallel to the magnetic field, causing the
geomagnetic radiation to almost vanish. Due to the electric field pulses appear with polarization components in the NS and $z$ plane. For a
field of 100 V/cm the extra electric field contributions are of the order of
the original pulse amplitude, which is to be expected since $E$ and $cB$ are of the same order (see Appendix \ref{app:polarization}). 

Following the same reasoning, in an electric field of 1000 V/cm, pulses could be produced that are an order of magnitude larger in amplitude
than the pulses in the absence of an electric field. Indeed, in the right panel of Fig.\ \ref{multi_az0}, such behavior is visible, but only for the thin lines, which represent a simulation in which
electric field effects are only taken into account in REAS2 and not in CORSIKA. When the CORSIKA electric field routine is switched on, the
pulse amplitudes in the NS and $z$ plane
drop by an order of magnitude. In the EW plane, the pulse amplitudes are even smaller than the pulse amplitude in 
absence of an electric field. The reason for this drop in pulse amplitude is the 
direction of motion of the shower electrons and positrons. In a strong field the charges are deflected strongly into the electric 
field direction. For inclined showers in a vertical electric field, this means that the particles only move into the direction of an
observer close to the shower axis for a much shorter part of their trajectories, and less radiation reaches this observer.

Instead, the particles that are deflected into the (vertical) electric field direction will radiate towards different 
locations on the ground, but these contributions spread out over a large area and will nowhere give emission of significant intensity.  

\begin{figure}[htp]
\centerline{\epsfig{file=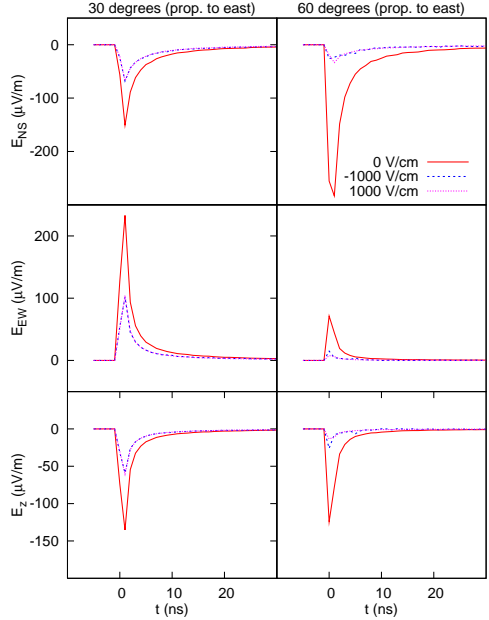 ,width=0.5\textwidth}}
\caption{Radio pulses for inclined showers of 10$^{16}$ eV propagating towards the east in the presence of electric fields of 100 V/cm (left panel) and 1000 V/cm (right panel). Pulses in the NS, EW and $z$
polarization are shown for an observers located 35~m north of the shower core. The electric field routine is switched on only in CORSIKA, not in REAS2.}
\label{multi_E0}
\end{figure}

The radio pulses for inclined showers propagating towards the east are shown in Fig.\ \ref{multi_az270}. For such showers the electric field contribution appears in the EW and
$z$ plane. Here also, the deviation in pulse height due to the electric field are of the order of the original pulse for a field of 100 V/cm. For a field of 1000 V/cm there is
again a large difference between the simulation with the CORSIKA electric field routine turned on or off. To isolate the effect of shower evolution, the electric field routine of
REAS2 is turned off in Fig.\ \ref{multi_E0}. The different lines correspond to showers that are simulated in CORSIKA 
with different field strengths, while in REAS2 the field is set
to zero for all three. The radio pulse of the shower with 30 degrees zenith angle has lost more than half of its amplitude, 
due to the deflection of particles. When the zenith
angle is larger the deflection is stronger and for 60 degrees the radiation has diminished by an order of magnitude.  
  
In Sec.~\ref{sec:effects} we have explained that the radio pulse calculation of REAS2 does not include the radiation from growing and decaying currents. In \cite{B09} we have
shown that CORSIKA simulations produce large increases in the number of electrons when the electric field exceeds the threshold field. 

The radio emission that is associated with
this pulse of runaway electrons is calculated by Gurevich et al.~\cite{G02}, Tierney et al.~\cite{T05} and recently by Dwyer et al.~\cite{D09}, and has characteristics that are very different from the geomagnetic pulse. 

First of all, the pulse is much stronger with values in the order of mV/m at kilometers distance for  a 10$^{17}$ eV shower passing through a typical thunderstorm field \cite{D09}.  

Second, since the mean propagation speed of the runaway electron avalanche is 0.89$c$ \cite{C06} the radio pulse from the avalanche is not relativistically beamed and also broader in time than the geomagnetic pulse. The time scale of runaway breakdown radio pulses is of the order of a microsecond, while the geomagnetic radio pulse is of the order of tens
of nanoseconds. In the latter case the pulse is shortened because the the radio waves and particles travel in the same direction, with almost the same speed. 

Because of these differences the radio pulse from the electron avalanche should be well distinguishable from the geomagnetic pulse.

\section{Discussion}
\label{sec:discussion}
The radio emission of air showers is driven by the deflection of electrons and positrons in the magnetic field. When an electric field 
is present, its contribution to
the total radiation can be approximated by comparing the perpendicular component of the electric force to the Lorentz force. Changes 
in radio pulse height due to an electric field are of the same order of the
original pulse height when $E_{\mathrm{\perp}}\sim cB$. For the geomagnetic field strength in central Europe of $B\sim 0.5$~G, this means an electric field of the order of 100 V/cm can alter the
radio pulse height significantly, while fields of the order of 1000 V/cm can dominate the emission mechanism. The geometry of the shower and the fields affects the various contributions. In a shower that propagates parallel to the electric field the
charges undergo only linear acceleration, for which the radiation field is suppressed by a factor $\gamma$. Since the bulk of the radiation is produced by particles with $\gamma
> 100$ for observers near the shower core and $\gamma>10$ at larger lateral distances \cite{H07}, the electric field contribution to the radio pulse is greatly 
suppressed in this case. In a shower that propagates parallel to the magnetic field the charges experience only a
small Lorentz force, leading to a small radio pulse. For such showers an electric field can have a relatively large influence.

\begin{figure}[htp]
\centerline{\epsfig{file=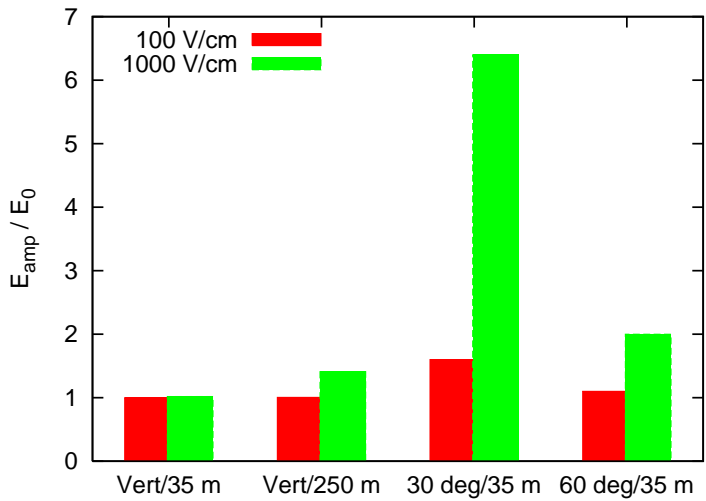 ,width=0.5\textwidth, angle=270}}
\caption{Ratio of heights of the pulses of showers with and without background electric fields. From left to right the ratios are given for the radio pulse of: 
a vertical shower observed at resp.~35~m and 250~m north of the shower core, and inclined showers with zenith angles of resp.~30 and 60 degrees propagating towards the north
observed at 35~m north of the shower core.    \label{extraplot}}
\end{figure}

We define the height of the radio pulse as the maximum value of the radiation field $E=|{\bf E}|$ of the pulse. Fig.~\ref{extraplot} shows the ratios of $E_{\mathrm{amp}}$, 
the pulse height of a pulse from a
shower moving through an electric field, and $E_{0}$, the pulse height of the same shower in absence of an electric field, for various cases. For a vertical shower and an
observer located 35~m north of the shower core, the ratio is near unity even for a field of 1000~V/cm. An observer located 250~m to the north measures a ratio of 1.4 if the
field strength is 1000~V/cm and almost unity for 100~V/cm. For inclined showers the ratios are larger, especially at 30 degrees zenith angle, because for a shower propagating
towards the north, the angle between the shower axis and the geomagnetic field is small, so the geosynchrotron emission in suppressed. The plot shows the results for showers
propagating towards the north.

The CORSIKA simulations that are used as input have a low energy cut-off of 0.5 MeV. In the absence of an electric field these low energy electrons can be safely ignored. It is shown in Figs. 24 and 25 of Huege et al. \cite{H07} that near the shower core the contribution of electrons and positrons with $\gamma < 10$ to the total radio emission is insignificant. At larger distances the relative contribution grows because the radiation beams of the highest energy particles have a smaller opening angle, but the bulk of the radiation is still produced by particles with $\gamma = 10 - 1000$.

For weak electric fields, the slow electrons can still be ignored, but their contribution could be larger when they are accelerated to runaway energies, allowing them to become relativistic and produce more low energy electrons by ionization. Due to elastic scattering the average position of the electrons produced in this process moves at a speed of 0.89$c$ \cite{C06}. Therefore, over a distance of 1 km the particles will have spread out over an area of 100 m trailing the shower front, given that the avalanches propagate in the same direction as the shower. In the general case, in which the shower axis is not aligned with the electric field direction, the area will be even larger. For a ground-based observer, a radio contribution from the avalanche will spread out over a time window of at least tens of microseconds and can be distinguished easily from the geosynchrotron pulse (which is tens of nanoseconds wide). Moreover, electrons spread out over an area of hundreds of meters will not radiate coherently in the MHz regime, in which geosychrotron pulses are observed.

In the simulations we have used homogeneous electric fields for the entire atmosphere, which is of course an unrealistic scenario. In fair weather conditions there exists a
background electric field that has a strength of $\sim 1.5$~V/cm at ground level and falls off rapidly with altitude. Such a field is too small to significantly influence the strength of
the radio pulse. Clouds can contain internal electric fields and the strength of these fields depends on the type of cloud. Most clouds have fields that are of the order of 10
V/cm or smaller. For such fields the electric force is an order of magnitude smaller than the Lorentz force, so radio pulse variations due to electric fields will be smaller
than 10\%. The shower particles at the shower maximum and just above it contribute most to the radio pulse. The reason for this is that the $\gamma^{-1}$ emission cone of 
the particles that are at higher altitude covers a larger ground area and that geosynchrotron emission is produced more efficiently in region of low density \cite{H07}. Clouds that do not extend up to the altitude of the shower maximum are not likely to influence the radio
emission.

Only a few types of clouds can contain fields that are large enough to have a significant influence on the radio emission. Nimbostratus clouds are known to support fields of the
order of 100~V/cm \cite{book}. These clouds typically have a base altitude of $\sim 2$~km and a thickness of 2--3~km. A shower that has its maximum inside such a 
cloud could emit a radio pulse that is influenced by the electric field. However, if the electric field is aligned vertically the shower has to be inclined for the effect to
be significant, and inclined showers typically have their maximum at higher altitudes. Nevertheless, it seems possible that under the right conditions air showers can emit amplified
radio pulses when moving through a nimbostratus cloud. In observations of radio emission in the presence of nimbostratus conditions no 
amplified radio pulses were found \cite{B07}. 

The largest electric fields are found inside thunderstorm clouds. These clouds can extend up to $\sim$10~km altitude and locally the field can have a strength of 1000--1500~V/cm.
The radio pulse of an air shower passing through a thunderstorm is likely to be influenced by electric field effects. Firstly, in regions with a field of $\sim 100$~V/cm the
radiation of the charges can significantly be enhanced or suppressed. Secondly, in regions with a field strength of $\sim 1000$~V/cm the charges are strongly deflected into the electric field
direction, resulting in a decrease of radiation for inclined showers. Thirdly, when the background field exceeds the threshold field, an electron runaway breakdown process can
occur that produces a current emitting a radio pulse that is strong enough to be detected at hundreds of kilometers distance.     

Amplification of the radio pulse is most likely to occur for showers passing through thunderstorms, but it is not impossible that under other weather conditions the electric field can
influence the pulse height. In most cases, the polarization properties of the radio pulse can give an indication that the pulse was not created by a pure geomagnetic mechanism.
For example, for showers propagating towards the south or the north the geomagnetic pulse is in the EW plane, while the electric field contribution shows up in the NS and $z$-plane.
Generally, for most shower geometries the geomagnetic polarization properties will differ from the electric field contribution (see Appendix \ref{app:polarization}).

This effect can in principle be used to derive electric field properties from polarization data of air shower measurements. Most radiation is created at the shower maximum and
before. By comparing measured polarization data to polarization properties that are expected for pure geomagnetic emission, the polarity of the electric field can be determined
as well as an estimate of the field strength.

For experiments that detect radio pulses from air showers it is important to know when a pulse has been influenced by an electric field, since it no longer can be used as a
measure for the shower energy. Such experiments should keep weather information, so that measurements that have been recorded during thunderstorms can be excluded from the general analysis.
A more sophisticated way of filtering out electrically influenced pulses is to analyze the polarization properties. Presently, polarization studies are not yet carried out in
enough detail to validate the polarization properties predicted by REAS2. When the polarization of showers of different geometries is measured and understood well enough,
pulses with anomalous polarization properties can be filtered out and analyzed separately.

\section{Conclusion}
\label{sec:conclusion}
Atmospheric electric fields affect air showers and their radio emission in a number of ways. The shower electrons and positrons are accelerated and deflected in the electric field which leads
to altered energy distributions. The radio emission due to this acceleration is of the same order of magnitude as the radiation from geomagnetic deflection for fields of the
order of
100~V/cm. When the field exceeds the threshold field, runaway electron breakdown may occur, adding a new generation of electrons to the shower. The current that is produced this
way can produce a strong radio pulse. The runaway breakdown process can in principle initiate lightning. In this case, the radio signal of the air shower will be followed by
radio emission from electrical processes inside the thunderstorm and, ultimately, a discharge. 

When radio pulses of air showers are produced by the geomagnetic mechanism only, they can be used as a measure for the energy of the primary particle \cite{huegenew}. This technique becomes
unreliable when the pulse height is affected by an atmospheric electric field.
\begin{itemize}
\item For most weather conditions, atmospheric electric fields are too small to significantly change the strength of the radio pulse of cosmic ray air showers.
\item The radio emission from air showers that pass through thunderstorms can be amplified or suppressed strongly. The radio pulse height for such showers is not a reliable measure
for the shower energy.
\item During other weather conditions the radio pulse may be influenced if there is an electric field present of the order of 100~V/cm at the location around the shower maximum.
Nimbostratus clouds are known to harbor such fields but their vertical extent is generally smaller than thunderstorm clouds, so 
the radio emission is only affected if the shower maximum occurs relatively deep in
the atmosphere. 
\item Pulses that have been influenced by an electric field generally show polarization properties different from pulses that are produced by a pure geomagnetic effect. Polarization measurements therefore contain information of the electric field strength and polarity at a region around the shower maximum.
\end{itemize}
In order to effectively filter out pulses that have been significantly affected by an electric field, a radio air shower experiment should keep weather information and do full
polarization measurements. 

Presently, the polarization of air shower radio pulses is being studied \cite{isar, codalema}. Such studies deepen our understanding of the emission mechanism and can provide a powerful filter against pulses that have been affected by electric fields.

\appendix

\section{Particle trajectory in electric and magnetic field}
\label{app:trajectory}
\begin{figure}[htp]
\centerline{\epsfig{file=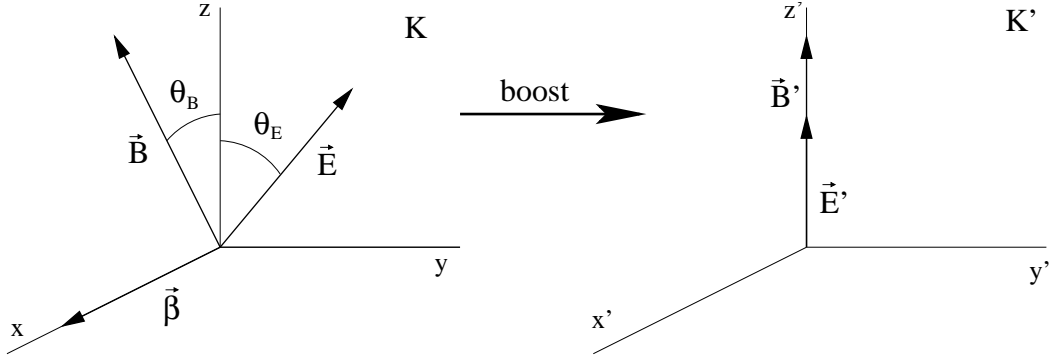 ,width=\textwidth}}
\caption{Coordinate frame $K$ contains a uniform $E$ field and a uniform $B$ field under an angles of respectively $\theta_E$ and
$\theta_B$ with the $z$-axis. In Lorentz-boosted frame $K^{\prime}$ the fields are aligned along the $z^{\prime}$-axis.}
\label{hacoord} 
\end{figure}
The trajectory of a particle with initial velocity ${\bf \beta}_i$ in an electromagnetic field is derived. The field consists
of a uniform electric field ${\bf E}$ that has an angle $\theta$ to a uniform magnetic field ${\bf B}$.
There exists a Lorentz boost ${\bf \beta}_b$ in the ${\bf E}\times{\bf B}$-direction, that transforms these two fields to aligned fields.
We define a frame $K$ in which the ${\bf E}$ and ${\bf B}$ field vectors are in $zy$ plane, so the boost is along the $x$-axis. We choose 
the $z$-axis in such a way that both the ${\bf E}^{\prime}$ and ${\bf B}^{\prime}$ field vector are aligned with the $z^{\prime}$-axis in 
the Lorentz boosted frame $K^{\prime}$. The angles of ${\bf E}$ and ${\bf B}$ with the $z$-axis are respectively $\theta_E$ and $\theta_B$,
and $\theta= \theta_E+\theta_B$. 

In the boosted frame $K^{\prime}$ the fields are (Eqn.~11.149 in Jackson, 1975):
\begin{equation}
{\bf E}^{\prime}=\gamma_{b}({\bf E}+{\bf \beta}_{b} \times {\bf B}) - \frac{\gamma^2_{b}}{\gamma_{b}+1}{\bf \beta}_{b}({\bf \beta}_{b} \cdot {\bf E}),
\end{equation}
\begin{equation}
{\bf B}^{\prime}=\gamma_{b}({\bf B}-{\bf \beta}_{b} \times {\bf E}) - \frac{\gamma^2_{b}}{\gamma_{b}+1}{\bf \beta}_{b}({\bf \beta}_{b} \cdot {\bf B}).
\end{equation}
A boost ${\bf \beta}_b$ in the ${\bf E}\times{\bf B}$-direction gives:
\begin{equation}
{\bf E}^{\prime}=\gamma_{b}(0, E_{y}-\beta_{b} B_{z}, E_{z}-\beta_{b} B_{y}),
\end{equation}
\begin{equation}
{\bf B}^{\prime}=\gamma_{b}(0, B_{y}+\beta_{b} E_{z}, B_{z}-\beta_{b} E_{y}).
\end{equation}
The components of the original fields can be found by setting the $y$-components of ${\bf E}^{\prime}$ and ${\bf B}^{\prime}$ to zero. Let $\tan
\theta_{B}=B_{y}/B_{z}$ and $\tan \theta_{E}=E_{y}/E_{z}$. Then:
\begin{equation}
\tan 2 \theta_{E} = \frac{\sin 2 \theta}{\cos 2 \theta + E^2/B^2},
\end{equation}
\begin{equation}
\tan 2 \theta_{B} = \frac{\sin 2 \theta}{\cos 2 \theta + B^2/E^2},
\end{equation}
and the needed boost is:
\begin{equation}
\beta_{b}^2=\tan \theta_{E} \tan \theta_{B}.
\label{neededboost}
\end{equation}
The boost is in the ${\bf E}\times{\bf B}$-direction. In frame $K^{\prime}$ the particle trajectory is found to be:
\begin{equation}
x^{\alpha}=(ct,x,y,z)=\left(\begin{array}{c}
C \sinh\rho(\phi-\phi_{1}) \\ A R \sin (\phi- \phi_{0}) \\ A R \cos (\phi-\phi_{0}) \\ C \cosh \rho (\phi-\phi_{1}) \end{array}\right),
\label{solutioninKprime}
\end{equation}
where $\phi=\omega \tau$, $\omega=e B^{\prime}/m c$, $R=c/\omega$, $\rho=E^{\prime}/B^{\prime}$ and $C=R \sqrt{1+A^2}/\rho$. The initial
conditions are given by $A$, $\phi_{0}$ and $\phi_{1}$, which can be related to the initial velocity ${\bf \beta}_{i}$. To do this the initial
velocity vector must first be transformed to the frame $K^{\prime}$. The $x$-component translates as:
\begin{equation}
\beta^{\prime}_{x,i}=\frac{\beta_{x,i}-\beta_{b}}{1-\beta_{x,i} \beta_{b}},
\end{equation}
while 
\begin{equation}
\beta^{\prime}_{y,i}=\frac{\beta_{y,i}}{\gamma_{b}(1-\beta_{x,i} \beta_{b})},
\end{equation}
and
\begin{equation} 
\beta^{\prime}_{z,i}=\frac{\beta_{z,i}}{\gamma_{b}(1-\beta_{x,i} \beta_{b})}.
\end{equation}
The initial conditions are now given by:
\begin{equation}
\tan \phi_{0} = \frac{\beta^{\prime}_{y,i}}{\beta^{\prime}_{x,i}},
\end{equation}
\begin{equation}
\phi_{1}=\frac{1}{2\rho} \ln \frac{1-\beta^{\prime}_{z,i}}{1+\beta^{\prime}_{z,i}},
\end{equation}
\begin{equation}
A=\pm\sqrt{\frac{\gamma^{\prime 2}}{\cosh^{2} \rho \phi_{1}} -1},
\end{equation}
where the sign of $A$ can be found by evaluating the derivative of the $x$-component of Eqn.~\ref{solutioninKprime}. This component must have the same sign
as $\beta^{\prime}_{x,i}$ for $\tau=0$. The solution (Eqn.~\ref{solutioninKprime}) can now be boosted back to $K$:
\begin{equation}
x^{\alpha}=\left( \begin{array}{c} 
C \gamma_{b} \sinh \rho (\phi-\phi_{1}) + A R \gamma_{b} \beta_{b} \sin (\phi - \phi_{0}) \\
C \gamma_{b} \beta_{b} \sinh \rho (\phi-\phi_{1}) + A R \gamma_{b} \sin (\phi - \phi_{0}) \\
A R \cos (\phi-\phi_{0}) \\
C \cosh \rho (\phi-\phi_{1}) \end{array} \right).
\label{location}
\end{equation}
This is the solution for a coordinate system in which ${\bf E}$ and ${\bf B}$ define the $yz$-plane, with the $z$-axis making an
angle $\theta_{E}$ with the electric field and an angle $\theta_{B}$ with the magnetic field. 

Differentation of Eqn.~\ref{location} to proper time $\tau$ gives:
\begin{equation}
U^{\alpha}=c\left(\begin{array}{c}
\gamma_{b} \sqrt{1+A^{2}} \cosh \rho (\phi-\phi_{1}) + A \gamma_{b} \beta_{b} \cos (\phi-\phi_{0}) \\
\gamma_{b} \beta_{b} \sqrt{1+A^{2}} \cosh \rho (\phi-\phi_{1}) + A \gamma_{b} \cos (\phi-\phi_{0}) \\
-A \sin(\phi-\phi_{0})\\
\sqrt{1+A^{2}} \sinh \rho (\phi-\phi_{1}) \end{array} \right),
\end{equation}
so
\begin{equation}
\gamma(\tau)=\frac{U^{0}}{c}=\gamma_{b} \sqrt{1+A^{2}} \cosh \rho (\phi-\phi_{1}) + A \gamma_{b} \beta_{b} \cos (\phi-\phi_{0}).
\end{equation}
The total energy of the particle can now be written as a function of the proper time:
\begin{equation}
E(\tau)=\gamma(\tau) m c^{2}.
\end{equation}

If the angle between the electric and magnetic field is greater than 90 degrees, a similar approach can be used. A boost can be applied in such a way that the
field will be anti-aligned. It is easy to verify that this boost is also in the ${\bf E} \times {\bf B}$-direction. The boost is still given by Eqn.~\ref{neededboost}.
Equation \ref{solutioninKprime} is still the right solution, but $\rho$ switches sign and will have a negative value.

In this derivation the $z$-axis was chosen in such a way that the transformed fields align along the $z^{\prime}$ axis. Beforehand, however, this
direction is unknown. The angles $\theta_{E}$ and $\theta_{B}$ must first be derived from $E$, $B$ and the angle $\theta$. From this, an angle can be calculated
between the $z$-axis used in this derivation and the `natural' $z$-axis of the sky. The solution (Eqn.~\ref{location}) must be rotated with this angle to find the
solution in sky coordinates.

The solution is given in proper time. In the REAS2 code this has to be translated to the time in the observer frame. The relation between proper and
lab time is:
\begin{equation}
t=\int^{\tau}_{0} \gamma(\tau^{\prime}) d\tau^{\prime},
\label{labtime}
\end{equation}
giving:
\begin{equation}
t=\frac{\gamma_{b} C}{c} \sinh \rho (\phi - \phi_{1}) + \frac{\gamma_{b} \beta_{b} A}{\omega} \sin (\phi-\phi_{0}).
\end{equation}

\section{Polarization of the radio pulse of a single particle}
\label{app:polarization}
We derive the polarization of the radio emission from a particle moving in the direction
\begin{equation}
\hat{n}=\left( \sin\theta\cos\phi,\sin\theta\sin\phi,-\cos\theta \right)
\end{equation}
where $\theta$ is the zenith angle of the particle ($\theta=0$ corresponds
to vertical downward motion) and $\phi$ is the azimuthal direction, defined such 
that $\phi=0^{\circ}$ corresponds to north, $\phi=90^{\circ}$ to west, etc.

The particle moves through a magnetic field
\begin{equation}
\label{eq:bfield}
{\bf B}=\left(\sin\eta,0,-\cos\eta \right)
\end{equation}
where we neglect a small azimuthal angle between the magnetic field
and the geographic north. There is a vertical electric field
\begin{equation}
{\bf E}=(0,0,E)
\end{equation}
present.
The particle will feel a Lorentz force
\begin{equation}
{\bf F}_B = qc \hat{n}\times{\bf B}
\end{equation}
where we use $v\sim c$ for a relativistic particle. The radiation field is
proportional to the force: ${\bf A}\propto\gamma^2 {\bf F}_B$, where $\gamma$
is the Lorentz factor of the particle. The perpendicular component of the electric force is given by
\begin{equation}
{\bf F}_E=q \left({\bf E}-({\bf E}\cdot\hat{n})\hat{n}\right)
\end{equation}
The radiation from the component of the electric force that is aligned with the direction of propagation, is suppressed by a factor $\gamma$ and is neglected here. For
the total radiation field we find:
\begin{equation}
{\bf A}\propto \left({\bf E}-({\bf E}\cdot\hat{n})\hat{n}\right) + c \left(
\hat{n}\times{\bf B} \right)
\end{equation}
or
\begin{equation}
{\bf A} \propto \left( \begin{array}{c}
E\cos\theta\sin\theta\cos\phi-cB\sin\theta\sin\phi\cos\eta\\
E\cos\theta\sin\theta\sin\phi+cB(\sin\theta\cos\phi\cos\eta-\cos\theta\sin\eta)\\
E\sin^2\theta-cB\sin\theta\sin\phi\sin\eta \end{array} \right)
\label{polarization}
\end{equation}

Two important properties of the radio emission from a particle in an electric and magnetic field are clear from this result. Firstly, the electric field gives a contribution to
the radiation that is of the same order as the magnetic radiation when $E\sim cB$. Secondly, the two contributions have different polarization properties, which makes it
possible to recognize pulses that were emitted in a region with a strong electric field.

For a complete air shower the radio pulse is found by integrating over the individual pulses of all particles which
are radiating in the direction of a certain observer. Since these particles do not share the same $\hat{n}$ and the
electric field is not expected to be homogeneous, our result cannot be generalized to complete air showers. Nevertheless, the changes in the polarization properties that are
introduced by the electric field for a single particle should be reflected in the radio pulse from the complete shower. Indeed, in Sec.\
\ref{sec:results} it is demonstrated that the polarization properties of complete showers can be explained with this simple approximation. 

The polarization of the radio pulse can therefore be used as an indicator for the presence of a strong electric field. Whenever the
polarization properties of an observed pulse differ from the expected properties based on a pure geosynchrotron pulse, it can be inferred
that another emission process is involved.

\end{document}